\documentclass[aps,prl,twocolumn,showpacs,nofootinbib]{revtex4}

\usepackage{graphicx}

\begin{document}


\title{Hall effect in doped Mott insulator: DMFT -- approximation}




\author{E.\ Z.\ Kuchinskii, N.\ A.\ Kuleeva, D.\ I.\ Khomskii$^*$, 
M.\ V.\ Sadovskii}

\affiliation{Institute for Electrophysics, RAS Ural Branch, Amundsen st. 106,
Ekaterinburg 620016, Russia}

\affiliation{$^*$II Physikalisches Institut, Universitaet zu Koeln, 
Zuelpicher Str. 77, 50937 Koeln,
Germany
}




\begin{abstract}
In the framework of dynamical mean field theory (DMFT) we analyze
Hall effect in doped Mott insulator as a parent cuprate superconductor.
We consider the partial filling (hole doping) of the lower Hubbard band and
calculate the dependence of the Hall coefficient and  Hall number on hole
doping, determining the critical concentration for sign change of the Hall
coefficient. Significant temperature dependence of the Hall effect is noted.
A good agreement is demonstrated with the concentration dependence of Hall 
number obtained in experiments in the normal state of YBCO. 
\end{abstract}


\maketitle


\section{Introduction}

In recent years much interest was attracted to experimental studies of Hall
effect at low temperatures in the normal state of high -- temperature 
superconductors (cuprates), which is achieved in very strong external magnetic 
fields \cite{Boeb,Tal1,Tal2}. The observed anomalies of Hall effect in these
experiments are usually attributed to Fermi surface reconstruction due to
formation of (antiferromagnetic) pseudogap and corresponding quantum critical
point \cite{PrTal}.

At the same time rather commonly accepted view is that cuprates are strongly
correlated systems and their metallic (superconducting) state is realized
as a result of doping of a parent Mott insulator, which can be described most
simply within the Hubbard model. However, there are almost no works devoted to
systematic studies of doping dependence of Hall effect in this model.
A common question here is what is determining the sign of the Hall coefficient?
At small hole doping of a parent insulator like La$_2$CuO$_4$ or underdoped YBCO,
it is obviously determined by hole concentration $\delta$. Then at what doping
level  shall we observe the sign change of Hall coefficient, when is there a
transition from a small hole Fermi surface to a large electron one? Solution of this
problem is quite important also for the general transport theory in strongly
correlated systems.

Rather general approach to study Hubbard model is the dynamic mean field theory
(DMFT) \cite{pruschke,georges96,Vollh10}. The aim of the present paper is
a systematic study of concentration and temperature dependence of Hall effect for
different doping levels in the lower Hubbard band within DMFT approach, as well
as comparison of theoretical results with experiments on YBCO \cite{Tal1}.
We shall see that surprisingly good agreement with experiment at quantitative
level can be achieved even for this elementary model.

\section{Basic relations}

In DMFT \cite{pruschke,georges96,Vollh10} the electron self -- energy in single --
particle Green's function $G({\bf p}\varepsilon)$ is local and independent
of momentum. Due to this locality both the usual and Hall conductivities are
completely determined by the spectral density:
\begin{equation}
A({\bf p}\varepsilon)=-\frac{1}{\pi}ImG^R({\bf p}\varepsilon).
\label{SpDens}
\end{equation}
In particular the static conductivity is given by:
\begin{equation}
\sigma_{xx}=\frac{\pi e^2}{2\hbar a}\int_{-\infty}^{\infty}d\varepsilon
\left( -\frac{df(\varepsilon)}{d\varepsilon} \right)
\sum_{{\bf p}\sigma}\left( \frac{\partial \varepsilon ({\bf p})}{\partial p_x}
\right) ^2
A^2({\bf p}\varepsilon),
\label{Gxx}
\end{equation}
while Hall conductivity \cite{pruschke}:
\begin{eqnarray}
\sigma^H_{xy}=\frac{2\pi^2e^3aH}{3\hbar^2}\int_{-\infty}^{\infty}d\varepsilon
\left( \frac{df(\varepsilon)}{d\varepsilon} \right)
\sum_{{\bf p}\sigma}\left( \frac{\partial \varepsilon ({\bf p})}{\partial p_x}
\right) ^2\times\nonumber\\
\times\frac{\partial^2 \varepsilon ({\bf p})}{\partial p_y^2}
A^3({\bf p}\varepsilon).
\label{Gxy}
\end{eqnarray}
Here $a$ is the lattice parameter, $\varepsilon ({\bf p})$ is the electronic dispersion,
$f(\varepsilon)$ is the Fermi distribution, and $H$ the magnetic field along
z -- axis. Thus the Hall coefficient is:
\begin{equation}
R_H=\frac{\sigma^H_{xy}}{H\sigma_{xx}^2}
\label{R_H}
\end{equation}
is also completely determined by the spectral density $A({\bf p}\varepsilon)$,
which will be calculated within DMFT \cite{pruschke,georges96,Vollh10}.
To solve an effective single -- impurity Anderson model in DMFT
we used numerical renormalization group (NRG) \cite{NRGrev}.

We performed rather extensive calculations of Hall effect for different models
of electronic spectrum. Below, keeping in mind comparison with the experimental data
on YBCO, we limit ourselves to the results obtained for two -- dimensional
tight -- binding model of electronic spectrum:
\begin{equation}
\varepsilon ({\bf p})=-2t(cos(p_xa)+cos(p_ya))-4t'cos(p_xa)cos(p_ya).
\label{SPtt'}
\end{equation}
In this model we shall consider her two cases:\\

(1) the model with electron transfers only between nearest neighbors ($t'=0$)
and the complete electron -- hole symmetry;\\

(2) the case of $t'/t=-0.4$, which is qualitatively corresponds to YBCO.

For other cuprates we should use different values of $t'/t$ ratio.

Further on, for two -- dimensional models, the static conductivity
will be measured in the units of universal two -- dimensional conductivity
$\sigma_0=e^2/\hbar$, while Hall conductivity -- in units of $e^3a^2H/\hbar^2$.
Correspondingly, the Hall coefficient (\ref{R_H}) is measured in units of
$a^2/e$.

\section{Results of calculations and comparison with experiment}

\begin{figure}
\includegraphics[clip=true,width=0.6\textwidth]{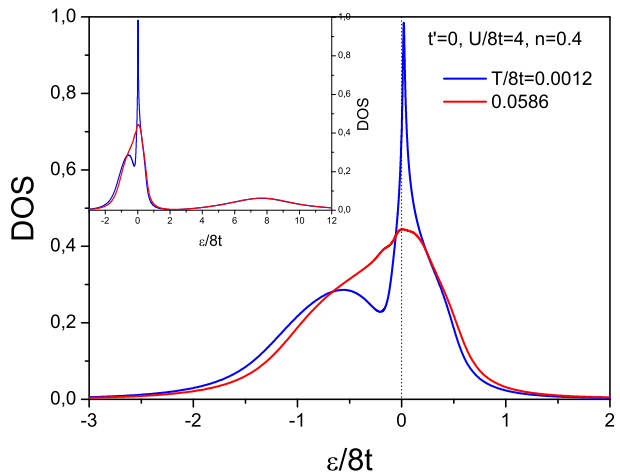}
\caption{Fig. 1. Density of states in doped Mott insulator for different
temperatures. Hubbard model parameters are shown in the Figure,
$8t$ -- initial bandwidth from (\ref{SPtt'}). At the insert we show the density
of states in wider energy interval including the upper Hubbard band.}
\label{fig1}
\end{figure}

For strongly correlated systems Hall coefficient is essentially dependent
on temperature. At low temperatures in these systems when treated in DMFT
approximation, besides upper and lower Hubbard bands also a narrow band forms close
to Fermi level forming the so called quasiparticle peak in the density of
states. In the hole doped Mott insulator (in the following we consider only the
hole doping) this peal lies close to the upper edge of the lower Hubbard band
(cf. Fig. \ref{fig1}). Thus, at low temperatures the Hall coefficient is
determined by the filling of this quasiparticle band.
At higher temperatures (of the order or higher than the width of quasiparticle
peak) the quasiparticle peak broadens and Hall coefficient is determined by
the filling of lower Hubbard band. Thus it is necessary to consider two
rather different temperature regimes for Hall effect.

\begin{figure}
\includegraphics[clip=true,width=0.5\textwidth]{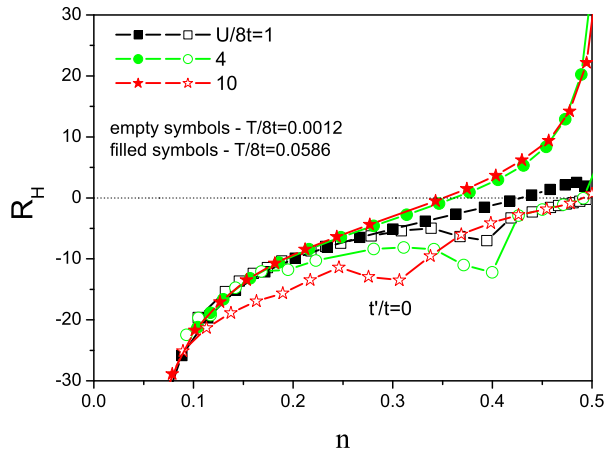}
\caption{Fig. 2. Dependence of Hall coefficient on band -- filling at low
(empty symbols) and high (filled symbols) temperatures for the model of two --
dimensional electron spectrum (\ref{SPtt'}) with transfers only between nearest
neighbors ($t'=0$).}
\label{fig2}
\end{figure}

In low temperature regime both the amplitude and width of quasiparticle peak depend
on band -- filling and temperature. Temperature growth leads to the brodening of
quasiparticle peak and some displacement of Fermi level below the maximum of
this peak (cf. Fig. \ref{fig1}). This may lead to a noticeable drop of Hall
coefficient, though further increase of temperature broaden the quasiparticle
peak and leads to the growth of this coefficient. Significant dependence of
quasiparticle peak on band -- filling in low temperature regime leads to the
regions of non monotonous dependence doping dependence of Hall coefficient
(cf. Fig. \ref{fig2}).

In high temperature regime the quasiparticle peak is strongly broadened and is practically
absent due to temperature. In this case, deeply in the hole doped Mott insulator
the Hall coefficient is in fact determined by filling of the lower Hubbard
band (the upper Hubbard band is significantly higher in energy and is practically
empty). In this situation, in the model with electron -- hole symmetry
($t'=0$) we can qualitatively estimate the band -- filling corresponding to sign
change of Hall coefficient as follows. Consider paramagnetic phase with
$n_{\uparrow}=n_{\downarrow}=n$, so that in the following $n$ denotes electron
density per single spin projection, while the total electron density is $2n$.
It is natural to assume that the sign change of Hall coefficient takes place
close to half -- filling of the lower Hubbard band $n_0\approx 1/2$. Consider
the states with ``upper'' spin projection, then the total number of states in
the lower Hubbard band is $1-n_{\downarrow}=1-n$. Then the band -- filling is
obtained as $n=n_{\uparrow}=n_0(1-n)\approx 1/2(1-n)$. Thus, for the
band -- filling corresponding to a sign change of the Hall coefficient we get
$n_c\approx 1/3$.

The same result is easily obtained also in Hubbard I approximation, where the
Green's function for spin up electrons is written as \cite{Khom}:
\begin{equation}
G^R_{\uparrow}(\varepsilon {\bf p})=\frac{1-n_{\downarrow}}{\varepsilon-
\varepsilon_-({\bf p})+i\delta}+
\frac{n_{\downarrow}}{\varepsilon-\varepsilon_+({\bf p})+i\delta}.
\label{HubbardI}
\end{equation}
where $\varepsilon_{\pm}(\bf p)$ is quasiparticle spectrum in upper and lower
Hubbard bands. We can see that in this approximation the number of states with
spin up projection in lower Hubbard band (first term in (\ref{HubbardI})) is
really $1-n_{\downarrow}$. During hole doping of Mott insulator the main
band -- filling goes into the lower Hubbard band, so that:
\begin{eqnarray}
&& n=n_{\uparrow}\approx\nonumber\\
&& \approx (1-n_{\downarrow})\int_{-\infty}^{\infty}d\varepsilon f(\varepsilon)
\left( -\frac{1}{\pi}Im\sum_{\bf p} \frac{1}{\varepsilon-\varepsilon_-({\bf p})+
i\delta}\right)=\nonumber\\
&& =(1-n)n_0.
\label{H1}
\end{eqnarray}
Then for half -- filled lower Hubbard band $n_0=1/2$ and the sign of Hall effect
(effective mass of quasiparticles) changes, so that we get $n=n_c=1/3$ again.

\begin{figure}
\includegraphics[clip=true,width=0.5\textwidth]{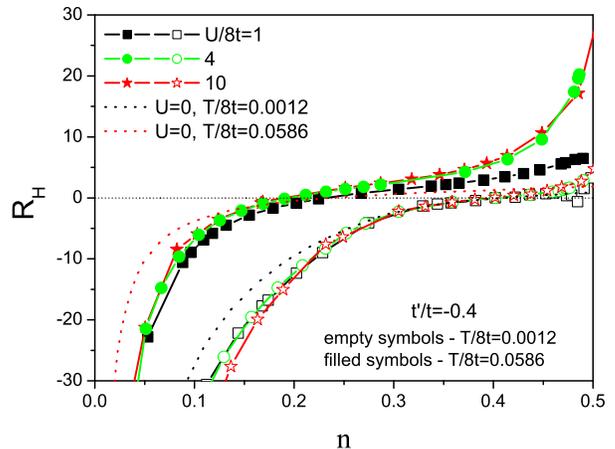}
\caption{Fig. 3. Dependence of Hall coefficient on band -- filling at low
(empty symbols) and high (filled symbols) temperatures for the model of two --
dimensional electron spectrum (\ref{SPtt'}) with transfers between nearest and
next -- nearest neighbors ($t'/t=-0.4$).}
\label{fig3}
\end{figure}

From Fig. \ref{fig2} it is easily seen that the high -- temperature behavior of
Hall coefficient in doped Mott insulator ($U/2D=4;10$)  in case of the complete
electron -- hole symmetry ($t'=0$)  fully supports this estimate. In case of
noticeable breaking of this symmetry the simple estimate does not work, as
even in the absence of correlations the sign change of Hall coefficient is
observed not at half -- filling (cf. Fig. \ref{fig3}).

It should be noted that the quasiparticle peak in the density of states is widened
and suppressed not only by temperature but also by disorder
\cite{dis_hubb_2008, GDMFT}, as well as by pseudogap fluctuations, which are
completely ignored in local DMFT approach \cite{DMFT+S,GDMFT}.
Thus, the range of applicability of the simple estimates made above for electron
-- hole symmetric case in reality may be significantly wider.

\begin{figure}
\includegraphics[clip=true,width=0.6\textwidth]{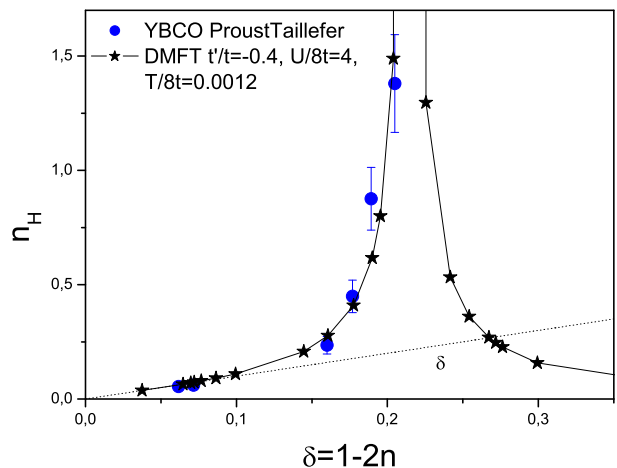}
\caption{Fig. 4. Dependence of Hall number $n_H$ on doping --- comparison with
experiment \cite{Tal1} on YBCO, $\delta=1-2n$ -- hole concentration.
Stars -- our calculations, blue circles -- experiment.}
\label{fig4}
\end{figure}

In Fig. \ref{fig4} we show the comparison of our calculations for the Hall number
(Hall concentration) $n_H=\frac{a^2}{|eR_H|}$ for typical model parameters
with experimental data for YBCO from Ref. \cite{Tal1}. We can see that even for
this, rather arbitrary, choice of parameters we obtain almost quantitative
agreement  with experiment, with no assumptions on Hall effect connection with
the Fermi surface reconstruction by pseudogap and closeness to corresponding
quantum critical point, which were used in Refs. \cite{Tal1,Tal2,PrTal}.
It is more or less obvious that similar data of Ref. \cite{Tal2} for NLSCO
can be interpreted within our model with appropriate change of parameters $t/t'$
and $U$. Thus it is quite possible that our interpretation of Hall effect in
cuprates based on the doping of lower Hubbard band of Mott insulator can be a
viable alternative  to the picture of quantum critical point.

It may be of great interest to make the detailed studies of the Hall effect in
the vicinity of critical concentration corresponding to the sign change of
Hall coefficient (divergence of Hall number). This can be done in the systems
(cuprates), where such sign change takes place under doping.

\section{Conclusions}

We have studied the behavior of the Hall coefficient in metallic phase appearing
due to hole doping of the lower Hubbard band of Mott insulator.
The change of sign of Hall effect in simplest (symmetric) case takes place 
close to doping $n_c=$1/3 per single spin projection or total electron density 
2/3 in the lower Hubbard band, correponding to hole doping $\delta=1-2n$=1/3, 
though in general case it depends rather strongly on the choice of model 
parameters. This concentration follows from simple qualitative estimates 
and is not related to more sophisticated factors,
such as change of topology of Fermi surface or quantum critical points.

More than satisfactory agreement of theoretical concentrations dependencies
obtained in the experiments on YBCO \cite{Tal1} shows, that our model may be a
reasonable alternative to the picture of Hall effect in the vicinity of
quantum critical point, related to closing pseudogap \cite{PrTal}.

The work of EZK, NAK and MVS was partly supported by RFBR Grant No. 20-02-00011.
The work of DIK was funded by DFG under project No. 277146847 - CRC 1238.

\end{document}